\documentclass{RNTI}
\usepackage{epsfig}
\begin{document}

\thispagestyle{empty}

\fancyhead{}
\fancyfoot{}
\fancyhead[LE]{
%
%
Expériences de classification de documents XML homogènes...
%
}
\fancyhead[RO]{
%
Despeyroux et al.
%
}
\fancyfoot[LE,RO]{RNTI - 1}

\begin{center}
{\Large\bf
%
Expériences de classification d'une collection de documents XML de
structure homogène 
%
}
~\\~\\
%
Thierry Despeyroux$^{*}$, Yves Lechevallier$^{*}$\\ 
Brigitte Trousse$^{**}$, Anne-Marie Vercoustre$^{*}$\\
%
~\\
%
%
$^{*}$Inria - Rocquencourt\\
B.P. 105 - 78153 Le Chesnay Cedex, France\\
$^{**}$Inria - Sophia Antipolis\\
B.P. 93 - 06902 Sophia Antipolis, France\\
email: Prénom.Nom@inria.fr\\
http://www-rocq.inria.fr/axis/

%
\end{center}
%
%
\begin{abstract}

Cet article présente différentes expériences de classification de 
documents XML de structure homogène, en vue 
d'expliquer et de valider  une présentation organisationnelle pré-existante.
Le problème concerne le choix des éléments et mots utilisés pour la 
classification et son impact sur la typologie induite.
Pour cela nous combinons une sélection 
structurelle basée sur la nature des éléments XML et une 
sélection linguistique basée sur un typage syntaxique des mots.
Nous illustrons ces principes sur la collection des rapports d'activité 
2003 des équipes de recherche de l'Inria en cherchant
des groupements d'équipes (Thèmes) à partir du contenu de 
différentes parties de ces rapports. Nous comparons nos premiers 
résultats avec les thèmes de recherche officiels  de l'Inria.

\end{abstract}
%

%
\section{Introduction}


Les documents XML sont maintenant incontournables et la classification
de ces documents est un domaine de recherche très actif, en
particulier pour définir des modèles de représentations de documents
qui étendent les modèles traditionnels en tenant compte de la
structure du texte (Yi and Dundaresan, 2000), (Denoyer and al.). Cela
revient souvent à considérer que les même mots apparaissant dans des
éléments XML différents sont en fait différents.  Ces approches sont
génériques, elles peuvent s'appliquer quelque soit la DTD, alors que
notre approche suppose une connaissance d'une sémantique implicite des
éléments pour les sélectionner.




Certaines
méthodes de classification réduisent les documents XML à leur partie
purement textuelle, sans prendre avantage de la structure qui
pourtant véhicule une information riche.
Nous nous intéressons à l'impact du choix des parties de documents
sélectionnées sur le résultat de la classification, l'idée étant que
ces différentes parties participent à différentes vues pouvant mener à
des classifications différentes. 
Nous pratiquons successivement deux
niveaux de sélection~: une sélection utilisant la
structure du document, puis une sélection
linguistique au niveau du texte précédemment sélectionné.  Nous
utilisons ensuite un algorithme de classification 
qui va
construire une partition des documents, affecter les documents à des
classes et  exhiber la liste des mots qui ont permis la classification.

Nos expériences utilisent la collection des rapports d'activité (RA) en
XML rédigés par les équipes de recherche de l'Inria. Nous
cherchons à regrouper les équipes de recherche travaillant dans des domaines proches,
et à comparer le résultat aux deux groupements officiels en thèmes de
recherche, le premier utilisé jusqu'en 2003 (Thèmes 2003), le second proposé en
2004 (Thèmes 2004).

\section{Le rapport d'activité de l'Inria}

L'Inria publie un rapport d'activité annuel
dont l'annexe scientifique est accessible sur le WEB.
La version 2003 de cette annexe peut être trouvée à l'adresse~:\\
http://www.inria.fr/rapportsactivite/RA2003/index.html. C'est un
ensemble de rapports en anglais produits par chaque équipe de recherche.
La version XML de ces rapports contient
146 fichiers, soit 229 000 lignes ou plus 14,8 Moctets de données.

La structure logique est
définie par une DTD. Si le style et le contenu sont libres, la forme
requiert certaines sections alors que d'autres sont optionnelles~:

{\tt <!ELEMENT raweb (accueil, moreinfo?, composition, presentation,}

{\tt  fondements?,domaine?, logiciels?, resultats,contrats?,}

{\tt  international?,diffusion?,biblio) >}

Les sections obligatoires contiennent la liste des membres de l'équipe,
une présentation des objectifs, les résultats nouveaux et les
publications. Les sections facultatives contiennent les pré-requis
(fondements), les domaines d'application, les logiciels produits ainsi
que les diverses collaborations.

Les équipes de recherche sont regroupées dans des {\it thèmes} dont le
rôle est de faciliter la présentation de l'activité globale de
l'institut et de faciliter leur évaluation.  Le nombre de
ces thèmes, et leur composition, vient de changer récemment en
relation avec les objectifs stratégiques de l'institut ce qui a motivé
notre étude de l'impact du changement des Thèmes 2003 en Thèmes
2004. La question est donc de savoir s'il existe un regroupement
naturel des équipes en fonction du rapport d'activité et de voir
quelles sont les parties du texte caractérisent au mieux les équipes
vis à vis des présentations qui sont faites.

\section{Méthodologie pour classer les documents XML}

Notre objectif est de regrouper automatiquement les équipes Inria en
{\it thèmes} de recherche, en s'appuyant sur leur rapports
d'activité. Nous supposons donc que les RA reflètent les domaines de
recherche des équipes et que certaines parties de ces rapports sont
plus spécifiques que d'autres dans la caractérisation des domaines de
recherche des équipes; par exemple, nous supposons que les conférences
et revues dans lesquels les chercheurs publient sont très
représentatives de leur domaine.

\subsection{Sélection structurelle et sélection du vocabulaire}

La première étape consiste à sélectionner les parties des RA
pertinentes pour cette tâche de classification. Cette extraction
utilise les outils décrits dans (Despeyroux, 2004), mais d'autres
outils standard XML sont utilisables si l'extraction ne requièrt
pas d'inférence.

Comme nous supposons que différentes parties du RA peuvent jouer un
rôle différent pour la classification des équipes, nous avons mené
cinq expériences utilisant des parties prédéfinies du RA. Le but est
d'identifier les parties les plus significatives.

\begin{enumerate}

\item Expérience  K-F: Mots-clefs attachés à l'élément {\it fondation}

\item Expérience  K-all:  Tous les Mots-clefs du document.

\item Expérience T-P: Plein texte de l'élément {\it présentation}

\item Expérience T-PF: Plein texte des éléments {\it présentation} et
{\it fondation}

\item Expérience T-C: Noms des conférences, workshops, congrès, etc.

\end{enumerate}

La seconde étape de prétraitement est la sélection automatique de
mots représentatifs des cinq expériences de l'étape précédente.
Les méthodes traditionnelles de sélection de mots sont basées sur
des approches statistiques, par exemple utilisant la fréquence des
mots, ou le gain d'information. Nous avons choisi une approche
basée sur l'analyse de la langue naturelle, qui ne peut
généralement pas être appliquée pour un volume de textes trop
important en raison du coût de calcul. Nous utilisons l'outil
TreeTagger développé à l'Institut de Linguistique Computationelle
de l'Université de Stuttgart (Schmid, 1994). TreeTagger marque les
mots d'un texte avec des annotations grammaticales (nom, verbe,
article, etc.) et transforme les mots en leur racine syntaxique
(lemmatisation).

Pour les expériences K-F et K-all (mots-clefs), nous conservons les noms, verbes et adjectifs, tandis que pour T-PF et T-P
(texte), nous ne gardons que les noms. Il y a une difficulté avec
les conférences qui sont écrites de façon très hétérogène par les
équipes: certaines utilisent le nom complet, d'autre l'acronyme
sous des formes variées (e.g. POPL'03, POPL03, POPL 2003). Pour
résoudre ce problème, nous avons construit manuellement une liste
normalisée de conférences et réalisé automatiquement la correspondance entre
l'intitulé trouvé dans le RA et la forme normalisée. Enfin, nous
avons éliminé dans toutes les cas tous les mots qui ne sont pas
utilisés au moins une fois par deux équipes. La table
\ref{table:datasize} donne la taille des données utilisées dans
chaque expérience.

\begin{table}
{\scriptsize
\begin{center}
\begin{tabular}{|l|r|r|r|r|}\hline
 Experiences  & number   &  extrated   & selected   &  voca-  \\
        & of projects   &  words   &  words  &  bulary  \\ \hline
K-F & 80 & 2234  &  1053 & 134 \\
K-all & 121 & 8671 &   6171 & 382 \\
T-P & 138 & 63711 &  16036 & 365 \\
T-PF & 139 & 320501 &  87416 & 809 \\
T-C & 131 & 10806 & 7915 & 887\\
 \hline
\end{tabular} \caption{\label{table:datasize}Taille des données dans les différentes expériences}
\end{center}
}
\end{table}

\subsection{Méthode de classification et évaluation externe}
L'objectif de cette troisième étape est de regrouper les documents
(rapport d'activité d'une équipe) en un ensemble de classes
disjointes à partir des vocabulaires issus des cinq expériences
décrites dans le paragraphe précédent.

Pour réaliser ce classement nous utilisons une méthode de
partitionnement, proposée dans (Celeux et al., 1989), où la
distance est basée sur les fréquences des mots du vocabulaire
choisi. Le principe de l'algorithme est proche de l'algorithme des
k-means. Comme pour la méthode des k-means nous représentons les
classes par des prototypes qui résument, au mieux, l'information
des documents appartenant à chacune des classes.

Plus précisément, si le vocabulaire choisi est constitué de $p$
mots alors chaque document $s$ est représenté par le vecteur
$x_s=(x_s^1,...,x_s^j,...,x_s^p)$ où $x_s^j$ est le nombre
d'occurrence du mot $j$ dans le document $s$, alors le prototype
$g$ d'une classe $U_i$ est représenté par
$g_i=(g_i^1,...,g_i^j,...,g_i^p)$ où $g_i^j$ est calculé par
$g_i^j=\sum_{s \in U_i}{x_s^j}$.

Le prototype de chaque classe $i$ étant fixé, chaque
document est alors affecté à la classe dont la proximité du prototype avec ce document est la plus petite. Dans nos
expériences la proximité est calculée par une distance entre les
distributions associées aux documents et aux prototypes (Celeux et
al., 1989).

L'{\bf évaluation de la qualité des classes} générées par la
méthode de classification est basée sur sa comparaison avec les
deux listes de thèmes utilisés par l'Inria. Il s'agit d'une
évaluation {\it externe} puisque ces deux listes de thèmes n'ont
pas été utilisées par l'algorithme de classification. Pour cette
évaluation externe deux mesures complémentaires ont été utilisées:
la {\it F-measure} et le {\it corrected rand}.

La {\bf F-measure} proposée by (Larsen and Aone, 1999) combine les
mesures de {\it précision} et {\it rappel} bien connues en
recherche d'information. Soient $n_{ik}$ le nombre des équipes de
recherche Inria ayant leurs rapports classés dans la
classe $U_i$ et affectés au thème $C_k$,
$n_{i.}$ le nombre de rapports de la classe
$U_i$, $n_{.k}$ le nombre d'équipes du thème $C_k$ 
et $n$ est le nombre d'équipes analysées. Alors la
F-measure $F(i,j)$ entre le groupe calculé $U_i$ et le groupe
a priori $C_k$ est égale à $(2.*R(i,j)*P(i,j)/(R(i,j)+P(i,j))$ où
$R(i,k)=n_{ik}/n_{i.}$ est le rappel et $P(i,k)=n_{ik}/n_{.k}$ est
la précision.

La F-measure entre la partition a priori $U$ en $K$ groupes et la
partition $P$ des équipes Inria obtenue par la méthode de
classification est égale à:
\begin{equation} \label{eq:F}
F=\sum_{k=1}^{K}{\frac{n_{.k}}{n}*\max_j (F(k,j))}
\end{equation}

L'index  {\bf corrected Rand} ($CR$) a été proposé par (Hubert and
Arabie (1985)) pour comparer deux partitions. Nous rappelons sa
définition. Soit $U = \{U_1, \dots, U_i, \dots, U_r\}$ et $P =
\{C_1, \dots, C_k, \dots, C_K\}$ deux partitions d'un même
ensemble de équipes ayant respectivement $r$ et $K$ classes.
L'index du {\it corrected Rand} est:

\begin{equation} \label{eq:adjustedrand} CR =
\frac{\sum_{i=1}^r\sum_{k=1}^K {{n_{ik}}\choose{2}} -
{{n}\choose{2}}^{-1}\sum_{i=1}^r{{n_{i.}}\choose{2}}\sum_{=1}^K{{n_{.}}\choose{2}}}
{\frac{1}{2}[\sum_{i=1}^r{{n_{i.}}\choose{2}}+\sum_{=1}^K{{n_{.}}\choose{2}}]
-{{n}\choose{2}}^{-1}\sum_{i=1}^r{{n_{i.}}\choose{2}}\sum_{k=1}^K{{n_{.}}\choose{2}}}
\end{equation}

\noindent où ${{n}\choose{2}}=\frac{n(n-1)}{2}$.

La valeur de CR $\in [0, 1]$.
Une valeur proche de 0 correspond à une partition aléatoire.

\section{Analyse des résultats}
La table \ref{table:externalvalid} résume les résultats obtenus pour
différentes sélections de mots et différents nombres de classes (4, 5,
9). Nous analysons d'abord les résultas pour  Thèmes 2003 et Thèmes
2004 séparément, puis entre eux.

Avec les Thèmes 2003 comme référence, les meilleurs résultats sont
obtenus de façon consistante pour les deux mesures et dans tous les
cas quand le nombre de classes est fixé à 4. Une exception est le
groupement en 9 classes quand à la fois le texte des éléments {\it
présentation} et {\it fondation} (T-PF-c) est utilisé.  Le meilleur
résultat de tous est obtenu avec 4 classes qui utilise le texte des
éléments {\it présentation} et {\it fondation} (T-PF-a).

Une analyse plus fine utilisant les sous-thèmes (omise par manque de
place) montre une bonne correspondance entre classes et sous-thèmes,
sauf pour les sous-thèmes 1c et 2a regroupés dans la classe 4, et le
thème 3a divisé entre les classes 1 et 4.

Pour Thèmes 2004, nous obtenons les meilleurs résultats pour 5 classes et en utilisant tous les mots clefs (K-all-b).

Dans les deux cas, les sections {\it fondation}  semblent représentatives des domaines de recherche mis en avant par les thèmes, que ce soit par le texte de ces sections ou les mots-clefs qui y sont attachés (pour les équipes qui en ont fourni).

On peut noter que les comparaisons sont généralement
meilleurs avec  Thèmes 2003 qu'àvec Thèmes 2004, à l'exception du
cas de 5 classes crées en utilisant tous les mots-clefs. Il y a peu de
différences quand la classification se fait à partir des noms de
conférence. Bien que décevant, ceci peut s'expliquer par le fait que
nous n'utilisons pas TreeTaggeur dans ce cas, résultant en
l'utilisation d'un vocabulaire plus large et hétérogène (voir table
\ref{table:datasize}).

On peut aussi noter que les deux mesures utilisées sont très cohérentes entre elles: une valeur élevée de la F-measure correspond à un bon rand index. 

\begin{table}
{\scriptsize 
\begin{tabular}{|l|r|r|r|r|r|r|r|}\hline
 Exp.  &  Number & F   & Rand   &  F   &  Rand &  F   &  Rand \\
 & clusters & themes2003  &  themes2003   &  subthemes  &  subthemes & themes2004  &  themes2004\\ \hline
K-F-a   & 4 & {\bf 0.53} & {\bf 0.14} & 0.38 & 0.09 & 0.46 & 0.11 \\
K-F-b   & 5 & 0.44 & 0.05 & 0.35 & 0.06 & 0.37 & 0.03  \\
K-F-c   & 9 & 0.42 & 0.10 & 0.37 & 0.08 & 0.43 & 0.12\\ \hline
K-all-a & 4 & 0.52 & 0.17 & 0.36 & 0.09 & 0.47 & 0.15\\
K-all-b & 5 & {\bf 0.53} & {\bf 0.17} & 0.37 & 0.10 & {\bf 0.54} & 0.22 \\
K-all-c & 9 & 0.46 & 0.13 & 0.40 & 0.12 & 0.38 & 0.10\\  \hline \hline
T-P-a   & 4 & {\bf 0.55} & 0.19 & 0.40 & 0.14 & 0.50 & 0.19 \\
T-P-b   & 5 & 0.45 & 0.11 & 0.42 & 0.12 & 0.47 & 0.15 \\
T-P-c   & 9 & 0.44 & 0.11 & 0.45 & 0.16 & 0.44 & 0.14 \\ \hline
T-PF-a  & 4 & {\bf 0.66} & {\bf 0.32} & 0.49 & 0.27 & 0.50 & 0.21\\
T-PF-b  & 5 & 0.56 & 0.22 & 0.43 & 0.18 & {\bf 0.51} & 0.20 \\
T-PF-c  & 9 & 0.48 & 0.22 & {\bf 0.55} & 0.29 & 0.46 & 0.19 \\ \hline
T-C-a   & 4 & {\bf 0.51} & 0.15 & 0.39 & 0.15 & 0.50 & 0.21 \\
T-C-b   & 5 & 0.44 & 0.18 & 0.45 & 0.24 & 0.47 & 0.17 \\
T-C-c   & 9 & 0.45 & 0.13 & 0.47 & 0.21 & 0.45 & 0.15 \\
 \hline
\end{tabular} \caption{\label{table:externalvalid}Résultats avec mesure de  validité externe}
}
\end{table}

\section{Conclusion}

Nous avons présenté une méthodologie pour la classification de
documents XML de structure homogène et son évaluation par rapport à
deux typologies existantes.
Nous pensons que l'approche peut être utilisée pour d'autres collections XML dont la sémantique de la DTD est connue, même informellement.

Les résultats montrent que la qualité de la classification dépend fortement des  parties de documents sélectionnées. Dans notre application, l'utilisation de la partie {\it fondation} des rapports donne de meilleurs résultats que l'utilisation des mots-clefs.  On peut tourner ces conclusions à l'envers, comme une indication qu'un meilleur choix des mots-clefs et de la partie {\it présentation} permettrait de mieux décrire nos domaines de recherche stratégiques.




{\bf Remerciements}: Les auteurs tiennent à remercier Mihai Jurca, ingénieur dans l'équipe AxIS, pour son aide dans le prétraitement des données.

\subsection*{Références}

\begin{list}{~}{\setlength{\labelsep}{0 cm}
		\setlength{\leftmargin}{0.6 cm}
		\setlength{\rightmargin}{0 cm}}

\item \hspace{-0.6 cm}  Celeux G., Diday E., Govaert G.,
Lechevallier Y., Ralambondrainy, H. (1989): {\em Classification
Automatique des Donn\'{e}es, Environnement statistique et
informatique}. Bordas, Paris.

\item \hspace{-0.6 cm} Denoyer L., Vittaut J.-N., Allinari P.,
Brunessaux S., Brunessaux S. (2003), Structured Multimedia
Document Classification, In {\em DocEng'03}, Grenoble, France, pp 153-160.

\item \hspace{-0.6 cm} Despeyroux Th. (2004), Practical Semantic
Analysis of Web Sites and Documents, In {\em Proceedings of the
$13^{th}$ World Wide Web Conference (WWW2004)}, New York City, pp 685-693.

\item \hspace{-0.6 cm} Guillaume D., Murtagh F. (2000), Clustering of
XML documents, {\em Computer Physics communications}, Vol. 127, pp
215-227.

\item \hspace{-0.6 cm} Hubert L., Arabie P. (1985), "Comparing
Partitions", {\em Journal of Classification}, Vol. 2, pp 193-218.

\item \hspace{-0.6 cm} Larsen B., Aone C. (1999), Fast and
effective text mining using linear-time document clustering, In
{\em Proceedings of the fifth ACM SIGKDD international conference
on Knowledge discovery and data mining}, pp 16-22.

\item \hspace{-0.6 cm} Schmid H. (1994), Probabilistic
Part-of-Speech Tagging Using Decision Trees,In {\em Proc. of the
International Conference on New Methods in Language Processing},
Manchester, UK, pp 44-49.

\item \hspace{-0.6 cm} Yi J., Sundaresan N. (2000), A
classifier for semi-structured documents, In {\em Proc. of the
$6^{th}$ International Conference on Knowledge Discovery and Data
mining}, pp 340-344.

\end{list} 

\subsection*{Summary}
This paper presents some experiments in clustering homogeneous XML
documents to validate an existing 
organisational structure.
Our approach integrates techniques for extracting knowledge from
documents with document clustering method.  We focus on the
feature selection used for clustering and its impact on the emerging
classification;  
We mix the selection of structured features with the selection of 
textual features that is
based on syntactic typing by means of a tagger.
We illustrate and evaluate this approach with a collection of XML
activity reports written by Inria research teams for year 2003. The
objective is to cluster teams into larger groups (Themes) and to 
compare the results with Inria official research themes.

\end{document}